\documentclass[%
 reprint,
 amsmath,amssymb,
 aps,pre
]{revtex4-1}
\usepackage{graphicx}
\usepackage{dcolumn}
\usepackage{bm}
\usepackage{hyperref}
\usepackage{dblfloatfix}
\usepackage{float}
\usepackage[T1]{fontenc}
\usepackage{cases}
\usepackage{xcolor}
\begin{document}

\title{The Hitchhiker model for Laplace diffusion processes in the cell environment}

\author{M. Hidalgo-Soria}
\author{E. Barkai}%
 \email{Eli.Barkai@biu.ac.il}
\affiliation{%
 Department of Physics, Institute of Nanotechnology and Advanced Materials, Bar-Ilan University, Ramat-Gan 5290002, Israel\\
}%

\date{\today}
\begin{abstract}
Aggregation and fragmentation of single molecules  in the cell environment
lead  to statistical laws of movement different from Brownian motion. Employing
a many body approach, we  elucidate how the widely observed exponential tails
in the particle spreading, i.e. the Laplace distribution and the modulations of the
diffusivities, are controlled by size fluctuations of single molecules and for 
different dependencies of the diffusion on the size of particles. By means of
numerical simulations Laplace distributions are obtained whether we track one
molecule or many molecules in parallel. Using a renewal process in the space of
sizes, we quantify to what extent the diffusivity varies significantly depending on
which tracking protocol is applied.\\

\end{abstract}

\pacs{Valid PACS appear here}
                             
\maketitle

\begin{figure}[tbp]
\begin{center}
\mbox{\resizebox*{5.cm}{7.5cm}{\includegraphics{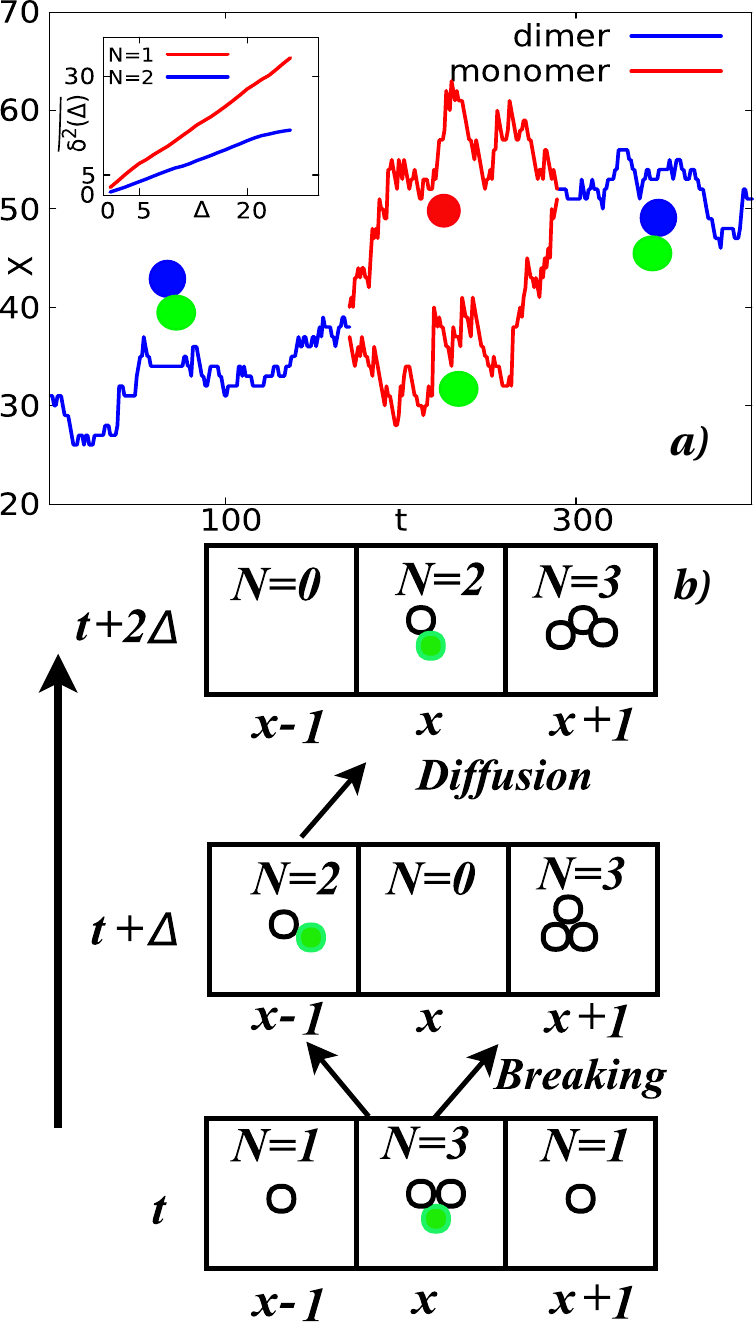}}} {}
\end{center}
  \vspace{-6mm}
\caption{{\protect\footnotesize {(Color online) \textit{a)}: Representative time series generated by the Hitchhiker model.  A  dimer, composed by a non tagged monomer and a fluorescent one, diffuses in space (blue curve) then it breaks into two monomers  which walk separately (red curves), there after they merge again (blue curves).  We show in the inset  the respective  time averaged MSD versus the lag time. \textit{b)}: Dynamics of the Hitchhiker model, at time $t$ we have certain configuration of molecules with different sizes. Then at time $t+\Delta$ a breaking event happens in the trimer at cell $x$, therefore a fluorescent monomer adds up to another one in the $x-1$ cell forming a dimer and the remaining two monomers merge with another one creating a trimer at cell $x+1$, leaving the site $x$ empty.  At time $t+2\Delta$ the dimer at cell $x-1$ jumps to the right.
}}}
\vspace{0mm}
  \label{fig:OLIGO}
\end{figure}

Imagine diffusive molecules in a  medium that can aggregate and break within the observation time of the experiment, see Fig.~\ref{fig:OLIGO}~\textit{a)}. Tracking these molecules individually will reveal that their diffusivities fluctuate. As the molecules are breaking and merging their sizes change, and this naturally leads to the speed up (small molecules) or slow  down (large molecules) of the stochastic dynamics. These processes are particularly important in the cell environment~\cite{wang2009,thomp2010,granick2012,wuite2013,berry2013,spako2017}. Since the diffusivity/size of tracked molecules is fluctuating we expect deviations from ordinary Brownian motion. In this letter, employing a many body approach called ``the Hitchhiker'', we address two main problems, how these deviations are induced and related to the widely reported Laplace packet spreading? and
 how the tagging protocol in single molecule experiments can  affect  the reported diffusivity?  

Firstly, in an increasing number of single molecule tracking experiments the diffusion of the tracer particles is shown to be linear, namely the mean square displacement ($MSD$) is  $\langle x^{2} \rangle = 2Dt$, with $D$ the diffusion coefficient. Hence according to Einstein's theory of Brownian motion one would expect that this normal behavior will come hand in hand with a Gaussian packet of spreading particles. Instead, in many cases as reported in ~\cite{wang2009,hapca2009,leptos2009,granick2012,peng2016,spako2017}, the tails of the density decay exponentially and this is modelled with the Laplace density $P(x,t)=\exp[-\mid x\mid / \langle D \rangle t] / \sqrt{4 \langle D \rangle t}$, with $\langle D \rangle$ the average diffusivity. Other experiments ~\cite{wuite2013,peng2016,spako2017} record the spectrum of $D$, and find that its distribution is an exponential one. As shown in ~\cite{chuby2014,sebastian2016,aki2016,chechk2017,Tyagi2017,Jain2018,vittoria2018,lan2018,LanG2018,grebenkov2019,jakub_2019} if we assume locally a Gaussian diffusive process  then averaging over the exponential distributed diffusivities  we get the Laplace probability density. Below, we start with a mathematically similar method, namely we ask what is the distribution of sizes of molecules that will induce a Laplace spreading?  This phenomenological method  shows that we may find Laplace distribution when either  the distribution of sizes is narrow or wide, depending on the interrelation between the local diffusion constant and  the size of the molecule
(Stokes-Einstein-Flory vs Arrhenius modeling, see below).  We then switch to a microscopical model, that allows us to check the hypothesis lade out in the phenomenological approach.  

As mentioned, the estimation of the diffusivity is important for the determination of the dynamics and reactions of molecules within the cell. However, the measured diffusivity may suffer a strong bias. For instance, for an \textit{ensemble} of diffusing identical particles in a medium, attaching to these a chromophore we will find a narrow distribution of diffusivities. Nevertheless, if the particle size is random, then the tendency of the chromophore is in most situations  to stick to larger size particles. Thus, we may encounter situations where the tagging protocol employed in single particle tracking experiments, favors the sampling of big size/slow particles. This also means that following all the particles in the \textit{ensemble} will lead to vastly different results if compared with the single particle tagging method. Below we quantify this behavior by means of the Hitchhiker model. Our work shows how the diffusivity reported in current single molecule experiments might be biased, due to the many body nature of the process, and also how to correct for this bias.     

{\em Phenomenological approach.} We assume that a polymer  has $N$ basic units, e.g. monomers \cite{doi1986}. We have in the system a large ensemble of these aggregates, so $N$ is random.  We first ask what is the spreading of  tracked particles or molecules  in this system, and this is given by:
\begin{eqnarray}\label{eq:SSX}
P(x,t)=\displaystyle \int _{0} ^{\infty} \frac{e^{- \frac{x^{2}}{4D(N)t} }}{\sqrt{4 \pi D(N) t}}P(N)dN,
\end{eqnarray}
This approach in its generality is sometimes called super-statistics ~\cite{beck2004}.  $P(N)$ is the distribution of the molecule sizes and  here $D(N)$ is the diffusion constant which depends on the size $N$. Our next question is phenomenological, given a diffusive law $D(N)$ what is the PDF $P(N)$ that yields the  observed Laplace distribution for $P(x,t)$?

A key feature of the process is the dependence of the diffusivity on $N$. We consider two different laws for $D(N)$, the Stokes-Einstein-Flory (SEF) and the Arrhenius one given by 
\begin{equation}\label{eq:phDN}
D(N)=\left\{
	\begin{array}{ll}
		\frac{k_{B}T}{6 \pi \eta b N^{\nu}}   & \mbox{SEF}, \\ \\ 
		D_{0}e^{-cN^{\tilde{\nu}}} & \mbox{Arrhenius}.  \\ \\
	\end{array}
\right.
\end{equation}
The SEF model uses a polymer chain size scaling, for which a macromolecule with a hydrodynamic radius $R$ and $N$ monomers satisfies: $R = bN^{\nu}$, where $\nu$ is the Flory exponent, and  $b$ the Kuhn length \cite{doi1986}. Typical values are: $\nu = 1$ the Rouse chain ~\cite{gennesI1976}, while Zimm chain gives $\nu= 3/5$ ~\cite{gennesII1976}. 
The Arrhenius model  has been used for describing the diffusion of proteins in polymer solutions, where an Arrhenius activation mechanism is known ~\cite{ullmann1985,holyst2013}. Here we have $D=D_{0} \exp[-E_{A}/k_{B}T]$ where $E_{A}$ is an activation energy. This activation energy depends on the size $N$ of the complex like $E_{A} = \epsilon bN^{\tilde{\nu}}$, with $\tilde{\nu}$ a scaling exponent and  $c = \epsilon  b / k_{B}T$ in Eq~.\eqref{eq:phDN} ~\cite{ullmann1985}. 

Using Eq.~\eqref{eq:SSX} and the SEF or the Arrhenius law with $\tilde{\nu} = 1$ Eq.~\eqref{eq:phDN}, in order to obtain the  Laplace law,  we have to employ (see Appendix  for details)  
\begin{numcases}{P(N)=}
  \frac{\nu k_{B}T}{6 \pi \eta b \langle D \rangle} N^{-\nu -1} e^{- \frac{k_{B}T}{6 \pi \eta b \langle D \rangle N^{\nu}}}, & SEF,\label{eq:phPN}\\
  \frac{cD_{0} }{\langle D \rangle} e^{-\big(\frac{D_{0}}{\langle D \rangle} e^{-cN} + cN \big)}, & Arrhenius.\,\,\,\,\,\,\,\ \label{eq:phPNA}
\end{numcases}

For the SEF model $P(N)$ has the form of a generalized inverse gamma distribution ~\cite{Mead2015}. This means that the distribution of sizes is fat tailed, in fact scale free in the sense that the mean of $N$ diverges when $\nu < 1$. In practice, in the model we study below, the power law tail must be cut-off due to finite size effects, still this law may capture the dynamics on some time scales. In the Arrhenius model $P(N)$ is the Gumbel density from extreme value statistics. Importantly, this type of distribution is peaked and narrow. Next we define our microscopical model.

{\em The Hitchhiker model.}  The Hitchhiker model was inspired by the experiments of Heller \textit{et al}. ~\cite{wuite2013} 
 and theoretical modelling of aggregation processes \cite{majum1998,raje2002}. Noteworthy the former deal with the diffusion of proteins on stretched DNA chains \textit{in vitro}, and they are depicted as a one dimensional system. The Hitchhiker model consists of an \textit{ensemble} of particles performing random walks on a lattice with size $L$ and with periodic boundary conditions. We start placing a  monomer ($N = 1$) on each lattice site. Given this,  at every time update one non-empty site is chosen randomly, then  either with probability $d(N)/[w+d(N)]$   we perform a diffusive step (see below) and the corresponding aggregation; or with probability $w/[d(N)+w]$ we perform a breaking event and its corresponding aggregation, see Fig.\ref{fig:OLIGO}~\textit{b)}. Here $d(N)$ and $w$ are respectively the rates of diffusion and breaking. Aggregates of monomers break into two, and then the remaining clusters are placed randomly at the immediate neighboring sites, leaving empty the site of breaking (see Fig.~\ref{fig:OLIGO}~\textit{b)}). In any case when  diffusion or breaking was held, if a neighboring site is already occupied, then particles meet and aggregation happens. When particles merge multi-meres are created, whose size is $N(t)$ ~\cite{oshanin1995,majum1998,raje2002}, then the diffusivity of the  particle $D\big[N (t)\big]$ is fluctuating in time. We have chosen binary breaking for the sake of simplicity but other breaking mechanisms like random scission or chipping  give similar results (see Appendix). $D$ and $d(N)$ are related by $D(N)\approx d(N)/2 \Delta$, with $\Delta=t_{i}-t_{i-1}$ the time increment (note that here the lattice spacing is set to one). The rate of diffusion $d(N)$ comprises the physical relation between $D$ and $N$ as following: $d(N)=1/N^{\nu}$ for the SEF model and $d(N)=\exp[-N^{\tilde{\nu}}]$ for the Arrhenius case. A key question is how will the SEF and Arrhenius approaches control the distribution $P(N)$ in equilibrium? and will this induce the Laplace spreading? 
 
\begin{figure}[tbp]
\begin{center}
\mbox{\resizebox*{5.5cm}{7.5cm}{\includegraphics{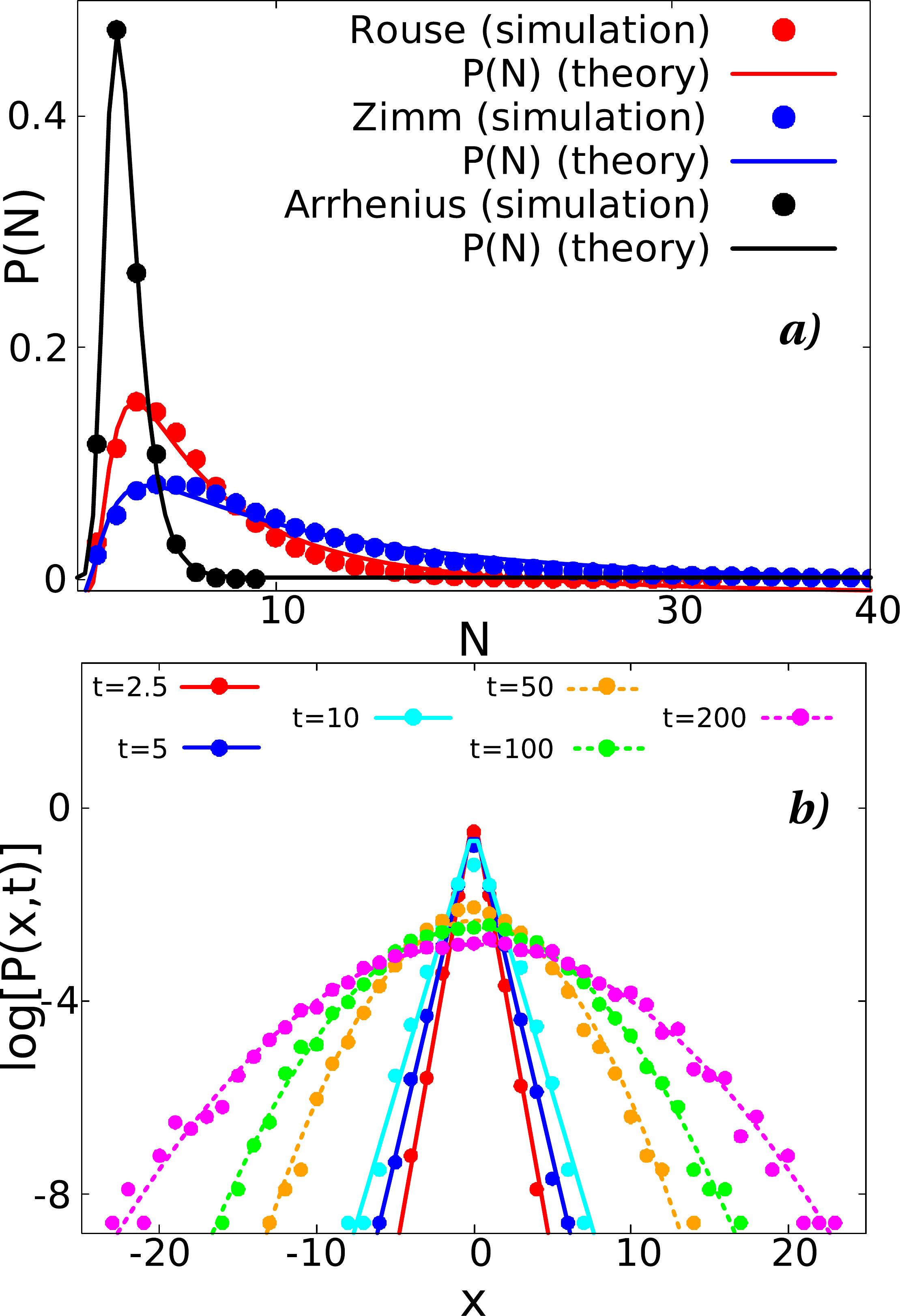}}} {} 
\end{center}
  \vspace{-6mm}
\caption{{\protect\footnotesize { (Color online)  \textit{a)}: Comparison between $P(N)$ obtained by simulations of the Hitchhiker model with the FT protocol and  the probability density Eq.~\eqref{eq:phPN}. We use three representative systems:  the Rouse model (red circles) and fitting of Eq.~\eqref{eq:phPN} with $\nu=1$ (red line), the Zimm model (blue circles) and  with $\nu=3/5$ (blue line), and Arrhenius model (black circles) and  theory  Eq.~\eqref{eq:phPNA} (black line).  \textit{b)}:  $P(x,t)$ in semi-log scale, obtained from the Hitchhiker model with Rouse dynamics. For short times   we show the comparison with their respective Laplace distribution  (solid lines). $P(x,t)$ for large times  are well described by Gaussian statistics  (dashed lines). For both cases the simulations were done for an \textit{ensemble} of $10000$ tracked molecules with  the FT method, $w=0.005$, $\Delta=1$ and in the steady state regime.
}}}
\vspace{0mm}
  \label{fig:pdfX}
\end{figure}

Fig.~\ref{fig:pdfX}~\textit{a)} shows clearly that modifying the microscopic law of diffusion has a strong impact on the distribution of $P(N)$. For SEF models i.e. with diffusion rates given by $d(N)=1/N$ for the Rouse model with $\nu=1$ and $d(N)=1/N^{\frac{3}{5}}$ for the Zimm model with $\nu=3/5$, we obtain visually broad distributions  of $P(N)$ well fitted by Eq.~\eqref{eq:phPN} while for the Arrhenius model, $d(N)=\exp[-N]$ with $\tilde{\nu}=1$, we find a very narrow distribution well fitted with Eq.~\eqref{eq:phPNA}. The sample average molecule size for the Zimm, Rouse and Arrhenius models satisfy the ordering $\langle N_{Z}\rangle=9.93 > \langle N_{R}\rangle=5.77 > \langle N_{A}\rangle=2.48$. Intuitively, the Arrhenius law causes large conglomerates to localize and hence this does not favor the creation of even bigger molecules (narrow distribution).

One of our main observations is that for three models of $D(N)$, i.e. Rouse, Zimm, and Arrhenius, the packet of particles exhibits a transition from Laplace distribution to a Gaussian behavior. In Fig.~\ref{fig:pdfX}~\textit{b)}  we show   $P(x,t)$ in semi-log scale for a system following the  Rouse model of diffusion rates and in the Appendix we show the corresponding for the Arrhenius law.  This transition was observed previously is some experiments ~\cite{wang2009} and using single particle Langevin dynamics with an added stochastic diffusivity ~\cite{chuby2014,chechk2017,vittoria2018}. For short times, relative to the breaking and merging rate, we observe different particles of different sizes, whose distribution is $P(N)$. Then to find the displacement we average the Gaussian propagator which depends on $D(N)$ over the respective distribution of sizes, which is exactly what we did already within the phenomenological approach Eq.~\eqref{eq:SSX}. We then get the Laplace law. However, for longer times each tracked single molecule, will fluctuate among many states, in each it will be attached to different number of particles. It follows that along a long trajectory we will average out the effect of fluctuating diffusivity and get in the long time limit Gaussian statistics.

Next we show the effects of the many body interaction on the measurement of diffusivity in single particle tracking experiments. Nowadays single molecule tracking techniques allow experimentalist to view many particles in parallel. Given these advances we explore theoretically  two tagging methods, one which follows all the complexes in the system, the full tagging (FT) method. And other where we have one light emitting particle which can be attached to different size complexes, the single molecule tagging method (SMT), see Fig.~\ref{fig:pdfzN}~\textit{a)}. We have found that in both tagging methods we get Laplace densities hence  in that sense it is universal (see details in Appendix).  We now ask how can we quantify the SMT dynamics and transform from one method to the other?
\begin{figure}[tbp]
\begin{center}
\mbox{\resizebox*{6.0cm}{7.5cm}{\includegraphics{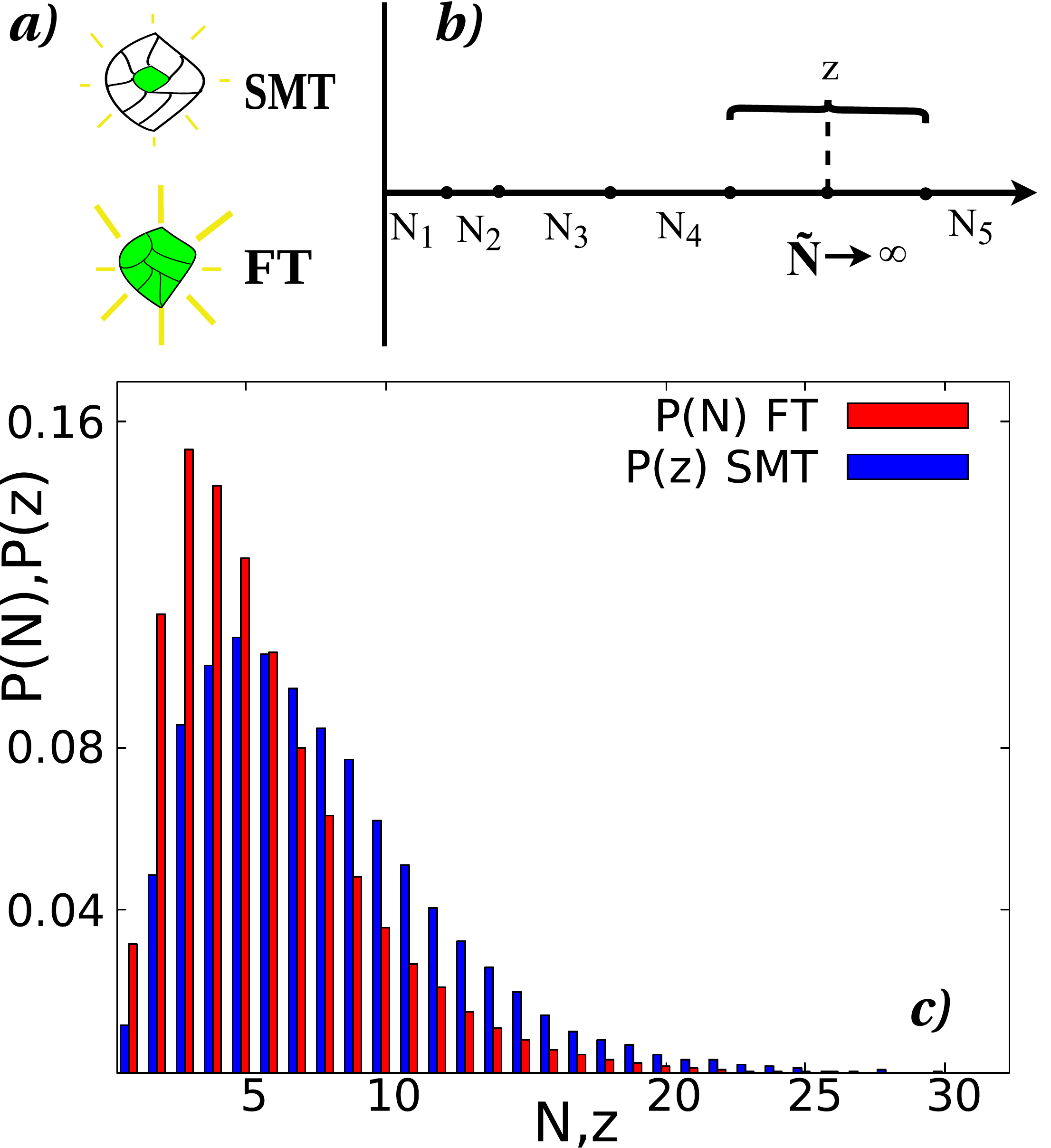}}} {} 
\end{center}
  \vspace{-6mm}
\caption{{\protect\footnotesize { (Color online)  \textit{a)} Graphic illustration of the SMT and FT protocols. \textit{b)} Schematic representation of the straddling size $z$ in the space of sizes for the SMT technique. \textit{c)} Comparison between the molecule size distribution $P(z)$  for the SMT method (blue boxes) and $P(N)$ for the FT protocol (red boxes) obtained by simulations of the Hitchhiker model with  the Rouse approach. The simulations were done using $w=0.005$, $\Delta=1$ and  $t= 10^{3}$ with an \textit{ensemble} of $10000$.  
}}}
\vspace{0mm}
  \label{fig:pdfzN}
\end{figure}

To quantify the difference in the  diffusivity arising from the usage of distinct  tagging methods we use  tools from renewal theory ~\cite{feller1960introduction,cox1962,wangwl2018}. In the SMT protocol, at the beginning of the experiment we  pick randomly one and only one monomer, and this is the tracked particle. At a given moment we have in the system complexes with different sizes:  $N_{1}, N_{2}, \ldots$,  etc. Placing all these complexes on the line, see Fig.~\ref{fig:pdfzN}~\textit{b)}, we then ask what is the distribution of the size  of the complex  on which the tagged monomer is residing (called $z$)? This mathematically is the same as defining some large $\tilde{N}$  (much larger than the average size) and asking where this $\tilde{N}$ will fall,  then the straddling size $z$ is defined by the interval around   $\tilde{N}$.   Repeating this procedure many times we can obtain the distribution of $z$.  In  ~\cite{feller1960introduction,cox1962,wangwl2018} it was found that it  satisfies
\begin{equation}\label{eq:strdNdist}
P(z)\sim \frac{N P(N)\vert_{N=z}}{\langle N \rangle}.
\end{equation}
Here $P(N)$ is the distribution of sizes of molecules in our system which we have investigated already, i.e, employing the FT method and  shown in Fig.~\ref{fig:pdfX}~\textit{a)}. Eq.~\eqref{eq:strdNdist} shows how larger molecules are more likely to be sampled, as we multiply $P(N)$ with $N$. To gain insights we now recover Eq.~\eqref{eq:strdNdist} using simple arguments.  Expressing  $P(z)$ as 
\begin{eqnarray}\label{eq:pzCL}
P(z)&=& \frac{\# \,\ of \,\ monomers \,\ in  \,\ complexes \,\ with \,\ size \,\ z}{total \,\ number \,\ of \,\ monomers \,\ in  \,\ the \,\ system}, \nonumber \\
& \simeq & \frac{z \cdot \# \,\ of \,\ complexes \,\ of \,\ size \,\ z}{\sum \limits_{i} \# \,\ of \,\ complexes \,\ of \,\ size \,\ i  \cdot \langle N \rangle}.
\end{eqnarray}
The last line of Eq.~\eqref{eq:pzCL} is the same as Eq.~\eqref{eq:strdNdist}, since by definition the empirical probability of the \# of complexes of size $z$ divided by the sum of \# of complexes of size $i$ is simply $P(N)_{N=z}$. Eq.~\eqref{eq:strdNdist} is soon to be used to predict the diffusivity in the SMT and compare it with the FT protocol.

Using  the Rouse model,  we proceeded to make simulations with the Hitchhiker model  employing the SMT protocol. The PDF of $z$ is shifted to the right compared with $P(N)$, namely large particles are sampled in agreement with Eq.~\eqref{eq:strdNdist}, see Fig.~\ref{fig:pdfzN}~\textit{c)}. As mentioned above, using the FT method we found $\langle N\rangle =5.77$. The peak (or the mode) of $P(N)$ is located at $N_{max}=3$. In the case of the SMT protocol we have a sample mean $\langle z\rangle=7.75$ and $z_{max}=5$, so $\langle z \rangle > \langle N \rangle$ as expected. Another interesting feature of $P(z)$ is that, in the large size regime, it has a fatter tail in comparison with the one of $P(N)$. 

Eq.~\eqref{eq:strdNdist} allows us to go from one measurement protocol to another, and to make predictions of the diffusivity and the spreading of packets. For example the diffusivity in equilibrium, in the single particle approach is $D(z)$ while  when we follow all the molecules we have $D(N)$. The general trend is that in the single molecule approach we sample large complexes, and hence the diffusion is slowed down compared with the full tagging approach since statistically $D_{SMT}(z) < D_{FT}(N)$. Employing Eq.~\eqref{eq:strdNdist} and the SEF model  we find that the ratio of the diffusivities meets
\begin{equation}\label{eq:RFTSM}
\frac{\langle D_{FT}\rangle}{\langle D_{SMT}\rangle}= \frac{\langle N \rangle}{\langle N^{1-\nu}\rangle}\Big\langle\frac{1}{N^{\nu}} \Big\rangle \geq 1.
\end{equation}
Here the averages on the right hand side are with respect to the distribution of sizes $P(N)$. This ratio is unity only if $P(N)$ is very narrow, i.e. it is delta peaked, or if $\nu = 0$ namely the diffusivity does not depend on size which is non-physical. In the Appendix we show  how Eq.~\eqref{eq:RFTSM} is satisfied for the Rouse dynamics, showing an example where $\langle D \rangle_{FT} / \langle D \rangle_{SMT} =1.44$, i.e. the diffusivity in the SMT protocol is diminished by $30\%$.  We have verified that also the SMT technique yields Laplace diffusion, the only effect is that the PDF of the particle spreading becomes narrower, since particles are slower, see Appendix. 

To summarize, employing the Hitchhiker model we showed that the mechanism that triggers the non Gaussianity is the aggregation between molecules and their sudden breaking.  The fluctuations in the molecule size  generates a diffusing diffusivity process, which exhibits non Gaussian distributions in $P(x,t)$,  such as single molecule experiments within the cell  like ~\cite{hapca2009,wang2009,leptos2009,thomp2010,granick2012,wuite2013,peng2016,spako2017}. We showed how the microscopic law of diffusion, i.e. SEF versus Arrhenius, strongly influences the distribution of sizes. In the Arrhenius case even a narrow distribution of molecule sizes can lead to a relatively large fluctuation in $D$. As expected, given $P(N)$ presented in Fig.~\ref{fig:pdfX}~\textit{a)},  in all cases we find in the short time regime Laplace distributions for $P(x,t)$, see Fig.~\ref{fig:pdfX}~\textit{b)}. 
The second main result  was that the protocol of tagging molecules matters. Employing the SMT protocol its average diffusivity is smaller around thirty per cent, compared with the diffusivity obtained via the FT protocol, Eq.~\eqref{eq:strdNdist} and Eq.~\eqref{eq:RFTSM} allow us to quantify this behavior.

As mentioned in the introduction  a major challenge is to determine the mechanism of the widely reported Laplace diffusion?  While we promoted the Hitchhiker approach, what can be said about other  microscopical models? Before answering this we would like to emphasize that our approach is based on experimental observations. Namely modern single molecule experiments, which follow the merging  of particles, can visualize the change of diffusivity and clearly correlate it to the aggregation/breaking. Or by measuring the intensity of the light, a proxy of size and correlating it to diffusivity, one can say that the mechanism we studied is clearly important in some experiments. Representative cases of such experiments are the diffusion of mRNA on  yeast cells ~\cite{thomp2010} or E. Colli cells ~\cite{spako2017}, and  of proteins in stretched DNA chains ~\cite{wuite2013}. Each one reports the respective correlation plot of the diffusivity and the intensity (which is proportional to the molecule size) or their distributions. 

However, we do not support the claim that this is the only approach, indeed it was shown that heterogeneity in the environment (without interactions) can lead to exponential tails, see ~\cite{kob2007,bb2019}. Also models of diffusion in narrow channels were promoted, see ~\citep{Li_2019}. In ~\cite{Seno2019} a model of a diffuser with a fluctuating size was considered, however they used a decoupled approach, particularly the distribution of sizes is not controlled by diffusion laws, while we showed the opposite trend: diffusion laws (SEF or Arrhenius) strongly determine the distribution of $N$, but in both cases yield Laplace spreading. Meanwhile, purely phenomenological models, \textit{e.g.} diffusing diffusivity models \cite{chuby2014,sebastian2016,aki2016,chechk2017,Tyagi2017,Jain2018,vittoria2018,lan2018,LanG2018,grebenkov2019,jakub_2019}, are clearly powerful as they allow for the estimation of reaction rates. We want to stress that our model recovers the experimental linking between diffusivity and the molecule size. This is achieved via the phenomenological diffusion law $D(N)$, and with the coupling of the diffusion rate with the aggregation dynamics. Furthermore, it gives qualitatively the same distribution of intensities/sizes as those found in ~\cite{thomp2010,wuite2013,heil2013,stracy2015}, and predicts that a Laplace density for $P(x,t)$ must emerge.

This work was supported by the Israel Science Foundation Grant No. 1898/17.
%

\appendix
\section{\label{sec:A1} Deduction of the distribution of sizes}
We are searching the distribution of sizes $P(N)$ which by means of Eq.~\eqref{eq:SSX} in the main text, satisfies the Laplace distribution $P(x,t)=\exp[-\mid x\mid/\langle D \rangle t]/\sqrt{4 \langle D \rangle t}$. By changing variables as $N \longrightarrow D$, now Eq.~\eqref{eq:SSX} is given by
\begin{eqnarray}\label{eq:SSX2}
\frac{e^{- \frac{ \mid X\mid }{\sqrt{ \langle D \rangle t}} }}{\sqrt{4 \langle D \rangle t}}=\displaystyle \int _{0} ^{\infty} \frac{e^{- \frac{X^{2}}{4Dt} }}{\sqrt{4 \pi D t}} P(D)  dD.
\end{eqnarray}
with $P(D)= (P(N)  \vert \frac{dN}{dD} \vert)  \vert _{N=N(D)}$ and $N(D)$ the inverse of $D=k_{B}T/6\pi \eta b N^{\nu}$ or $D=D_{0}\exp[-cN^{\tilde{\nu}}]$. As mentioned in the introduction the distribution of diffusivities must be exponential in order to obtain the Laplace law~\cite{chuby2014,chechk2017}. For the specific diffusion model the change of variables $D \longrightarrow N$ defines $P(N)$. In the Stokes-Einstein-Flory diffusivity model $D=k_{B}T/6\pi \eta b N^{\nu}$ the molecule size distribution is given in Eq.~\eqref{eq:phPN} in the main text. 

For the Arrhenius model we have $D=D_{0}\exp[-cN^{\tilde{\nu}}]$. The corresponding molecule size distribution for the case $\nu=1$ is reported in Eq.~\eqref{eq:phPNA} in the main text. 
\section{\label{sec:A2}Dynamics of the Hitchhiker model}
{\em Diffusion}. In the Hitchhiker model depending on their size the molecules have different probabilities of walking, such that the probability of hopping decreases with its respective size. Let $\lbrace X_{t} (N)\rbrace_{t \in \mathbb{Z}^{+}}$ be the position of a random walker with size $N$ at time $t$. As mentioned in the main text we employ a  size dependent diffusion rate $d(N)$, which is related with the diffusion coefficient $D$ see Eq.~\eqref{eq:DdN},  defined in each case by  $D=k_{B}T/6\pi \eta b N^{\nu}$ or $D=D_{0}\exp[-cN^{\tilde{\nu}}]$. For the Rouse model we have $d(N)=1/N$, in the Zimm model $d(N)=1/N^{\frac{3}{5}}$ and for the Arrhenius case $d(N)=e^{-N}$. Thus the corresponding transition probabilities from site $i$ to site $j$ in one step of time are given by
\begin{equation}\label{eq:TPRWm}
 P(X_{t}(N)=j \mid X_{t-1}(N)=i)=\left\{
	\begin{array}{ll}
		\frac{d(N)}{2}   & \mbox{if } j = i+1, \\ \\ 
		\frac{d(N)}{2} & \mbox{if } j = i-1, \\ \\
		1-d(N) & \mbox{if } j =i, \\ \\
		0 & \mbox{otherwise}.
	\end{array}
\right.
\end{equation}
For a molecule with size  $N$ the first/second row  in Eq.~\eqref{eq:TPRWm} defines the probability  to give a step to the right/left on the lattice. The third entry represents the probability  of remaining at the same site. Thus the displacement of a random walker with size $N$, is defined by $\Delta X_{t}(N)= X_{t}(N)-X_{t-1}(N)$, such that $\Delta X_{t}(N)\in \lbrace -1,0,1 \rbrace$. We consider two different sorts of interactions between particles: breaking and aggregation. 

{\em Breaking}. We assume a spontaneous binary breaking of molecules, see Fig.~\ref{fig:OLIGO} in the main text. Namely, if a molecule is composed of an even number of monomers it breaks into two equal parts of size $N_{i}/2$. When a molecule is composed by an odd number of monomers, it splits into two parts $(N_{i}-1)/ 2$ and $N_{i} - \frac{N_{i}-1}{2}$. In both cases the remaining clusters are placed randomly at the immediate neighboring sites, leaving empty the site of breaking. The rate of breaking is $w$ (see more details below in simulations methods).  

{\em Aggregation}. In the two cases of breaking, aggregation happens when the  remaining parts are placed randomly and add up with the molecules at their respective neighboring sites $j\in \lbrace i-1, i+1 \rbrace$, leaving the site $i$ empty. For the diffusion of particles at site $i$, the corresponding aggregation takes place  when the molecule jumps and adds up to the molecule at $i+1$ or at $i-1$, see trajectories in Fig.~\ref{fig:OLIGO} in the main text. 

The variance of a single displacement in the Hitchhiker model, which is defined by the displacements  $\Delta X_{t}(N)\in \lbrace -1,0,1 \rbrace$ and the transition probabilities Eq.~\eqref{eq:TPRWm}, is equal to 
\begin{equation}\label{eq:VarH}
\mathbb{E}[\Delta X_{t}^{2}(N)]=d(N).
\end{equation}
Substituting  Eq.~\eqref{eq:VarH} in $\langle x^{2} \rangle=2Dt$, the diffusion coefficient $D$ and the diffusion rate $d(N)$  in the Hitchhiker model are related by 
\begin{equation}\label{eq:DdN}
D=\frac{\mathbb{E}[\Delta X_{t}^{2}(N)]}{2 \Delta},
\end{equation}
with $\Delta =t_{i} - t_{i-1}$, here we used lattice spacing equal to one. The diffusion constant of the particle  is given by Eq.~\eqref{eq:DdN} ties the probability of choosing the molecule  at a given Monte Carlo step. The latter probability is  a constant in equilibrium.  Importantly it does not depend on the specific size of the molecule. Hence we have for a molecule $D(N) \sim d(N)$. In all of the cases (SEF or Arrhenius) the corresponding parameters in $D$ are set to one i.e. $k_{B}T / 6\pi \eta b=1$, $D_{0}=1$ and $c=1$.

\section{\label{sec:A3}Simulations}
The  simulations  of the Hitchhiker model were made by the following algorithm.  At the initial time  every site in the lattice is occupied  by one monomer ($N_{i}=1$), given this  at every update one non-empty site is chosen randomly, then   either with probability $d(N)/[w+d(N)]$   we do diffusion and the corresponding aggregation; or with probability $w/[d(N)+w]$ we perform a breaking event and its corresponding aggregation. We consider that after $M$ updates  a Monte Carlo step is achieved. The value of $M$ is defined by the average number on non-empty sites in the lattice, $1000$ for the Rouse Model, $500$ for the Zimm and $2000$ for the Arrhenius one. 

We use a fixed  rate of breaking $w=0.005$ and a lattice with $6000$ sites. And the rate of diffusion $d(N)$  depends on the size of each diffusing molecule via the Rouse, Zimm (Stokes-Einstein $D=k_{B}T/6\pi \eta b N^{\nu}$)  or Arrhenius $D=D_{0}\exp[-cN^{\tilde{\nu}}]$. The distributions of  $P(x,t)$ for all the cases were obtained within  the steady state regime for the molecule size distribution. This means that first we relax the system, letting it reach equilibrium.  
\section{\label{sec:A1}Distribution of displacements for the Arrhenius model}
\begin{figure}[tbp]
\begin{center}
\mbox{\resizebox*{5.5cm}{4.0cm}{\includegraphics{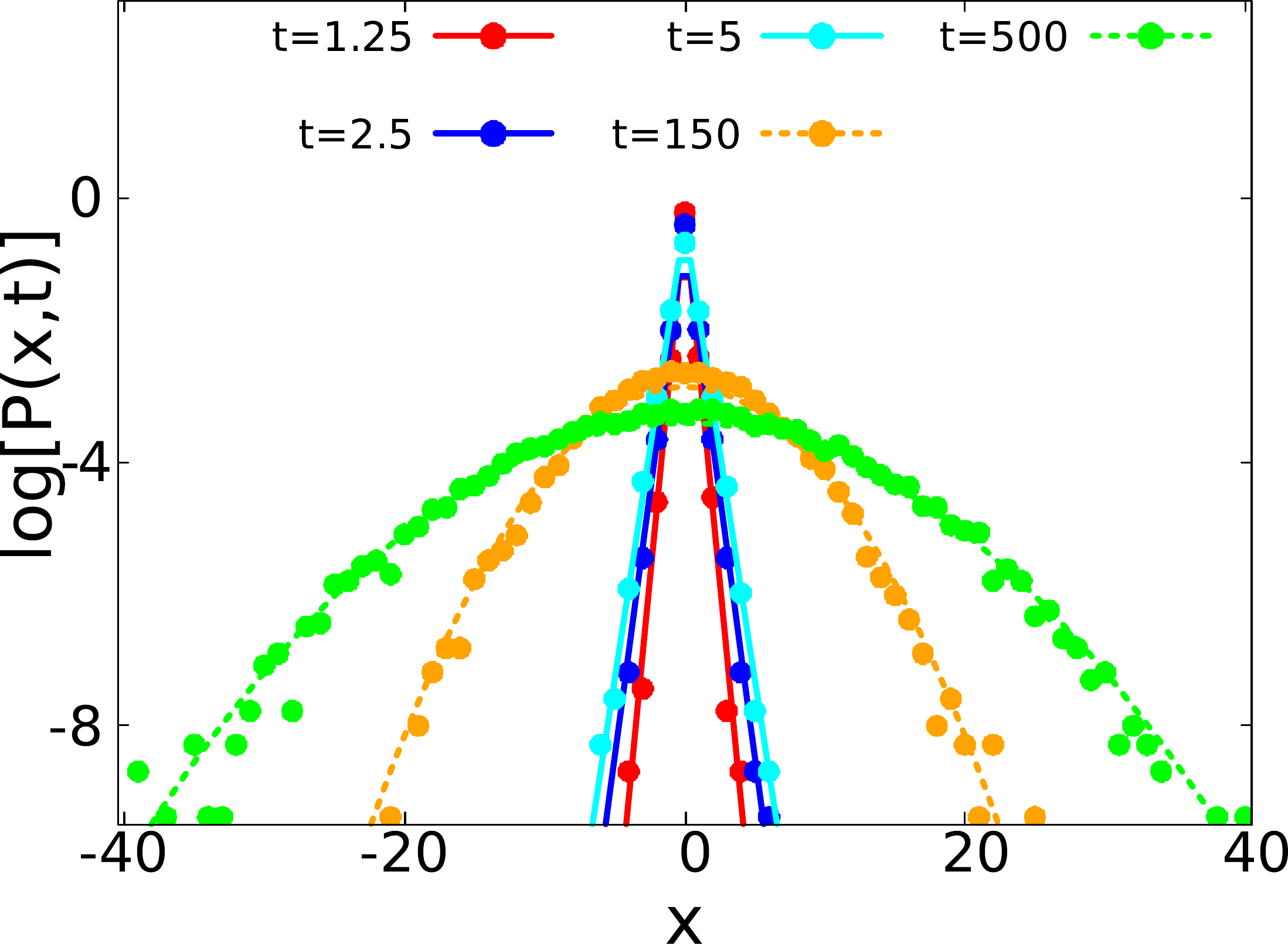}}} {} 
\end{center}
  \vspace{-6mm}
\caption{{\protect\footnotesize { (Color online) $P(x,t)$ in semi-log scale, obtained from the Hitchhiker model with Arrhenius diffusion rates. For short times   we show the comparison with their respective Laplace distribution  (solid lines). $P(x,t)$ for large times  are well described by Gaussian statistics  (dashed lines).the simulations were done for an \textit{ensemble} of $10000$ tracked molecules with  $w=0.005$, $\Delta=1$ and in the steady state regime. The particles were tracked by the FT method.   
}}}
\vspace{0mm}
  \label{fig:pdfxA}
\end{figure}

In Fig.~\ref{fig:pdfxA} we show the as in the case of Rouse dynamics in the main text, the distribution of displacements obtained by simulations of the Hitchhiker model with Arrhenius diffusion rates. As we can see for short times the molecule spreading is well fitted with the Laplace distribution. On the other hand, in the long time limit $P(x,t)$ follows Gaussian statistics. 

\section{\label{sec:A1}Comparison between analytical formula of $P(z)$ and simulations}

In  Fig.~\ref{fig:pdfzN}  we observe that  $P(z)$ (blue boxes) agrees with the analytical formula Eq.~\eqref{eq:strdNdist} extracted by the simulation data using the FT method (green boxes).

\begin{figure}[tbp]
\begin{center}
\mbox{\resizebox*{5.5cm}{4.0cm}{\includegraphics{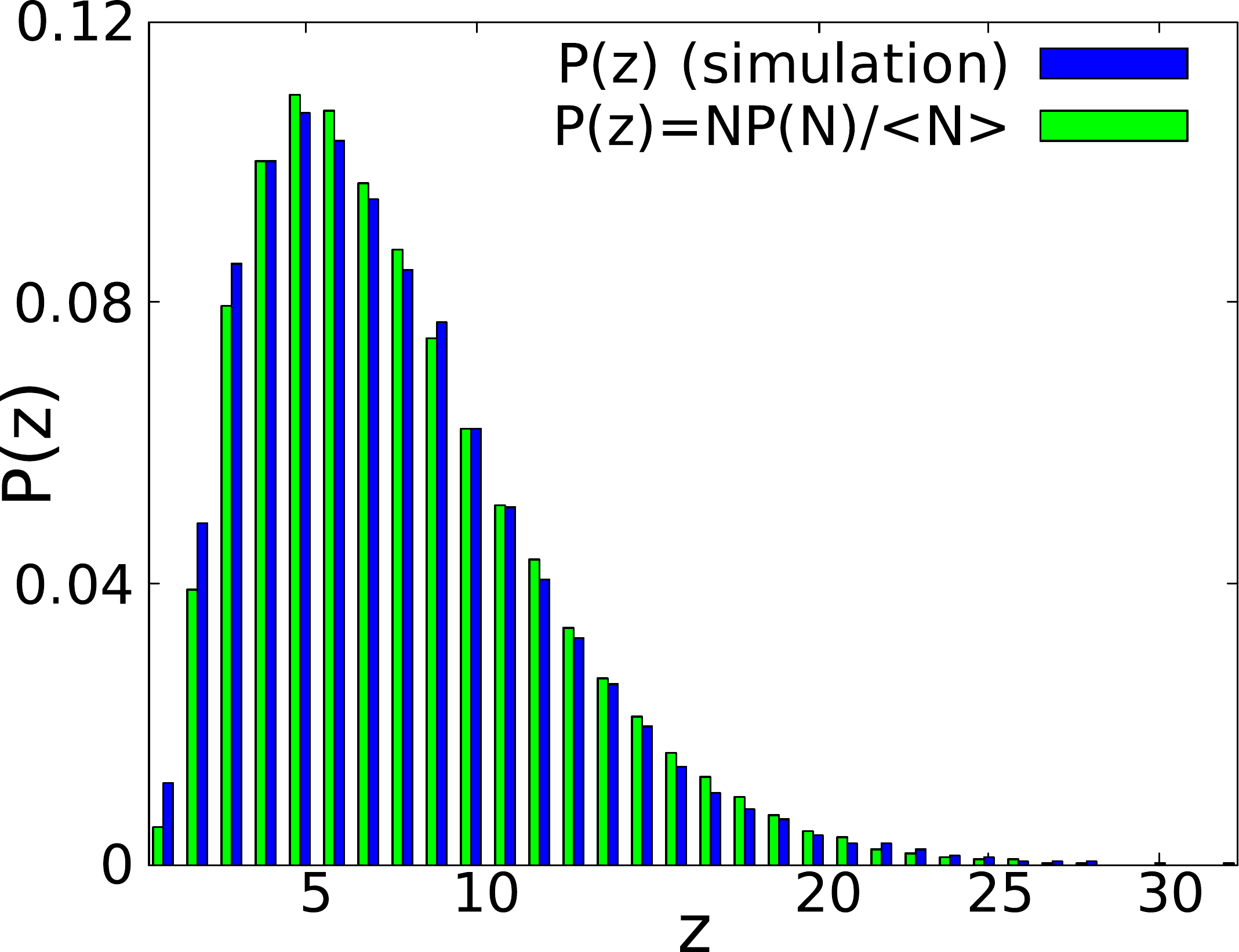}}} {} 
\end{center}
  \vspace{-6mm}
\caption{{\protect\footnotesize { (Color online) Comparison between $P(z)$ obtained by simulations (blue boxes) and $P(z)$ given by Eq.~\eqref{eq:strdNdist} (green boxes) employing the data of the simulations for the FT approach. Clearly Eq.~\eqref{eq:strdNdist}  well describes the data and with it we may obtain statistical properties of $D(z)$, see main text. The simulations were done using $w=0.005$, $\Delta=1$ and   $t= 10^{3}$ with an \textit{ensemble} of $10000$.  
}}}
\vspace{0mm}
  \label{fig:pdfzN}
\end{figure}

\section{\label{sec:A4}Average diffusivity in the FT and SMT tracking protocols} 
Following the Stokes-Einstein-Flory theory employing the diffusion rate $d(N)=1/N^{\nu}$, by Eq.~\eqref{eq:DdN} the diffusion coefficient is $D(N)\sim1/(2N^{\nu} \Delta)$. In this way  the ratio between average diffusivities in the FT and SMT methods is given by 
\begin{eqnarray}\label{eq:MDFT}
\frac{\langle D \rangle_{FT}}{\langle D \rangle_{SMT} } =  \frac{\langle d(N) \rangle_{FT}}{\langle d(N)\rangle_{SMT}}, 
\end{eqnarray}

and hence follows 
\begin{eqnarray}\label{eq:MDSM}
\frac{\langle D \rangle_{FT}}{\langle D \rangle_{SMT}}&=& \frac{\displaystyle \int \limits _{0} ^{\infty} d(N)p(N)dN}{\Bigg (\displaystyle \int \limits _{0} ^{\infty} \frac{N p(N)}{\langle N \rangle} d(N) dN \Bigg) \Bigg\vert_{N=z}} \\ \nonumber 
&=& \frac{\langle D_{FT}\rangle}{\langle D_{SMT}\rangle}= \frac{\langle N \rangle}{\langle N^{1-\nu}\rangle}\Big\langle\frac{1}{N^{\nu}} \Big\rangle.
\end{eqnarray}

\section{\label{sec:A5}Average diffusivity in the Rouse model} 
\begin{figure}[tbp]
\begin{center}
\mbox{\resizebox*{5.5cm}{4.0cm}{\includegraphics{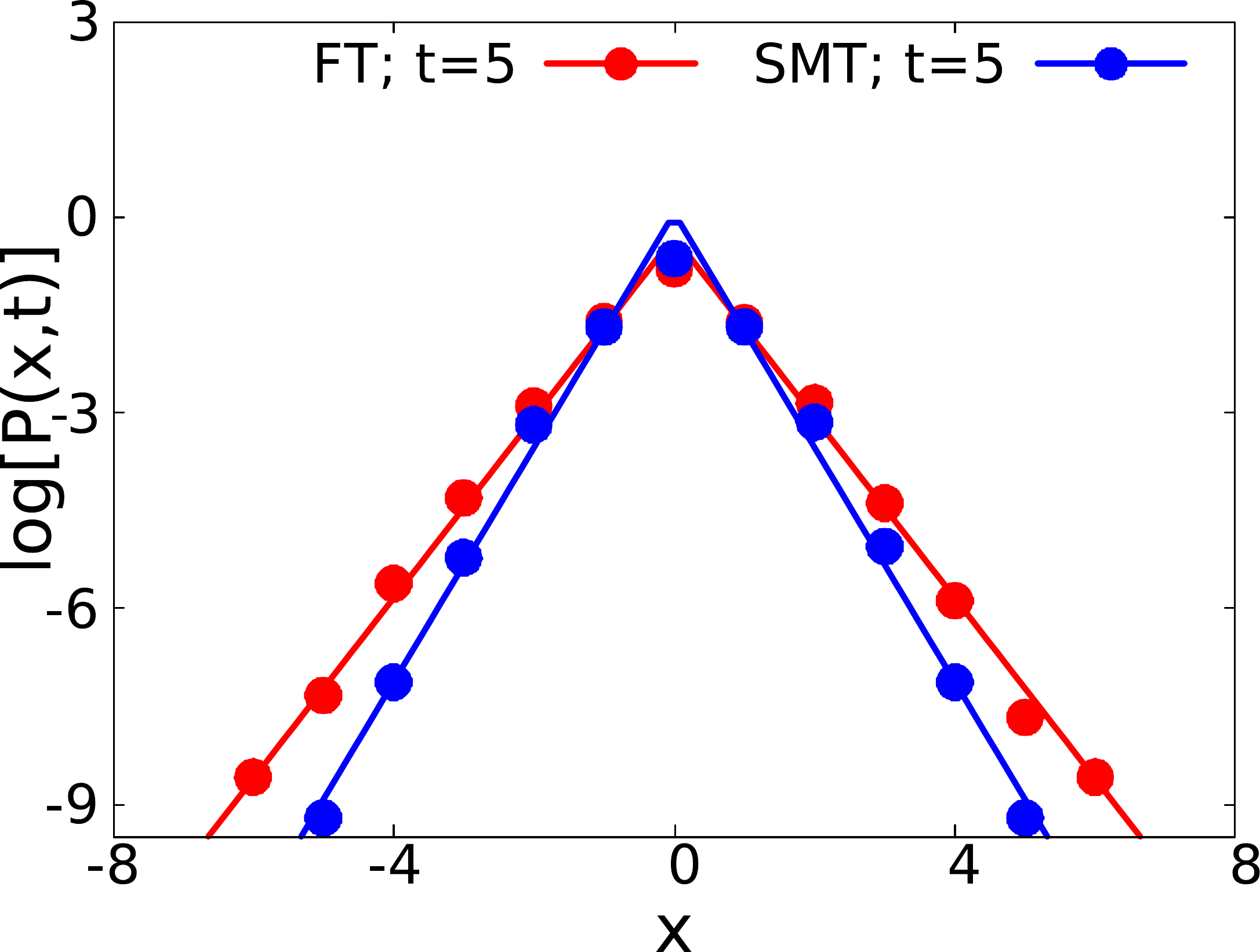}}} {} 
\end{center}
  \vspace{-6mm}
\caption{{\protect\footnotesize { (Color online)  For both tracking protocols we show  $P(x,t)$ in semi-log scale obtained  using the Rouse Model: the FT method (red circles) and  the SMT protocol (blue circles).  The  Laplace distribution is shown in solid color lines. In each case the measurements were done for $t=5$, $w=0.005$, $\Delta=1$ and  we tagged $10000$ molecules. Clearly the spread of the packet of particles using the SMT approach is much narrower, since single molecules are statistically favoring large complexes which are slowed down. 
}}}
\vspace{0mm}
  \label{fig:pdfXFTSMR}
\end{figure}
We corroborate this difference in the diffusivities for the Rouse model using simulations of the Hitchhiker model and  within the Laplace regime for $P(x,t)$,  in Fig.~\ref{fig:pdfXFTSMR}   we show the difference in the particle spreading generated by $D_{SMT}(z) < D_{FT}(N)$. When the SMT tagging method  is used  the maximum length in the displacements (blue circles) reached by the tracked particles is lower than the one obtained with the FT protocol (red circles). In each case we show in solid color lines the corresponding fitting with the Laplace distribution. 
From the data of $X_{t}$ we computed the ratio between diffusivities by the variance of each data set since $\langle x^{2} \rangle _{FT}= 2 \langle D \rangle_{FT} t$ and  $\langle x^{2} \rangle _{SMT}= 2 \langle D \rangle_{SMT} t$, having a value of $\langle D\rangle_{FT}/ \langle D\rangle_{SMT}=1.4550$. 

\section{\label{sec:A9}Other breaking mechanisms} 

\subsection{\label{sec:A91}Random scission}  
Instead of choosing for our model an equal binary breaking mechanism  here we present the distribution for molecule sizes $P(N)$ and displacements $P(x,t)$ for the Hitchhiker model, but for a random fission mechanism. For this case a breaking event happens at a constant rate $w$, although now  the cluster of particles with size $N_{i}$ is divided into two random parts $N_{i}-F$ and $F$. With $F$ a discrete uniform random variable, such that $F\in[1,N_{i}-1]$. In this way a monomer or a bigger sub-aggregate (less than $N-1$) can be ripped out from the cluster. The remaining two parts are placed randomly at the neighboring sites, leaving empty the site of breaking. The corresponding aggregation takes place when the mentioned fractions are add up at the neighboring sites $\lbrace i-1,i+1 \rbrace$. 
\begin{figure}[tbp]
\begin{center}
\mbox{\resizebox*{5.5cm}{4.0cm}{\includegraphics{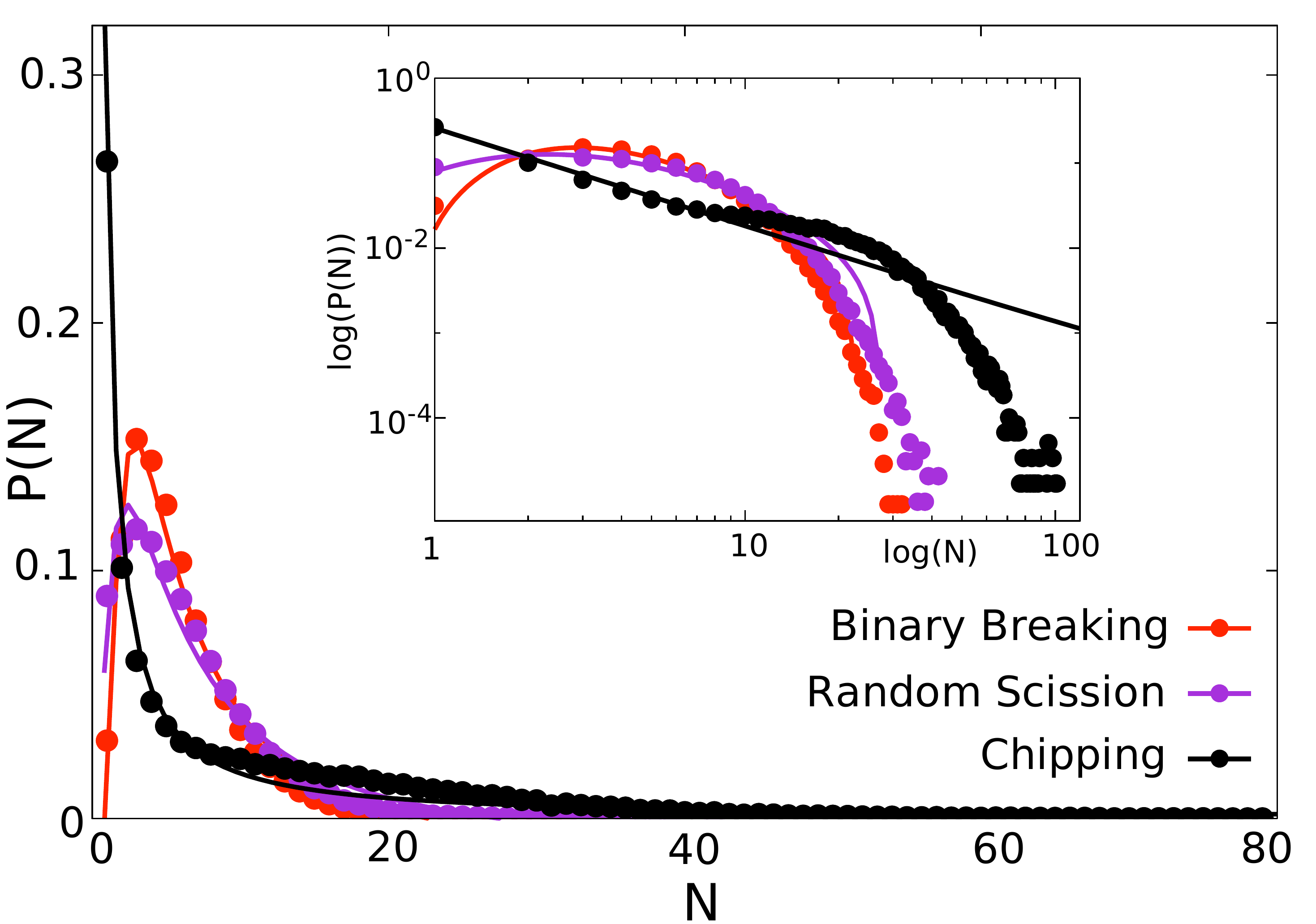}}} {} 
\mbox{\resizebox*{5.5cm}{4.0cm}{\includegraphics{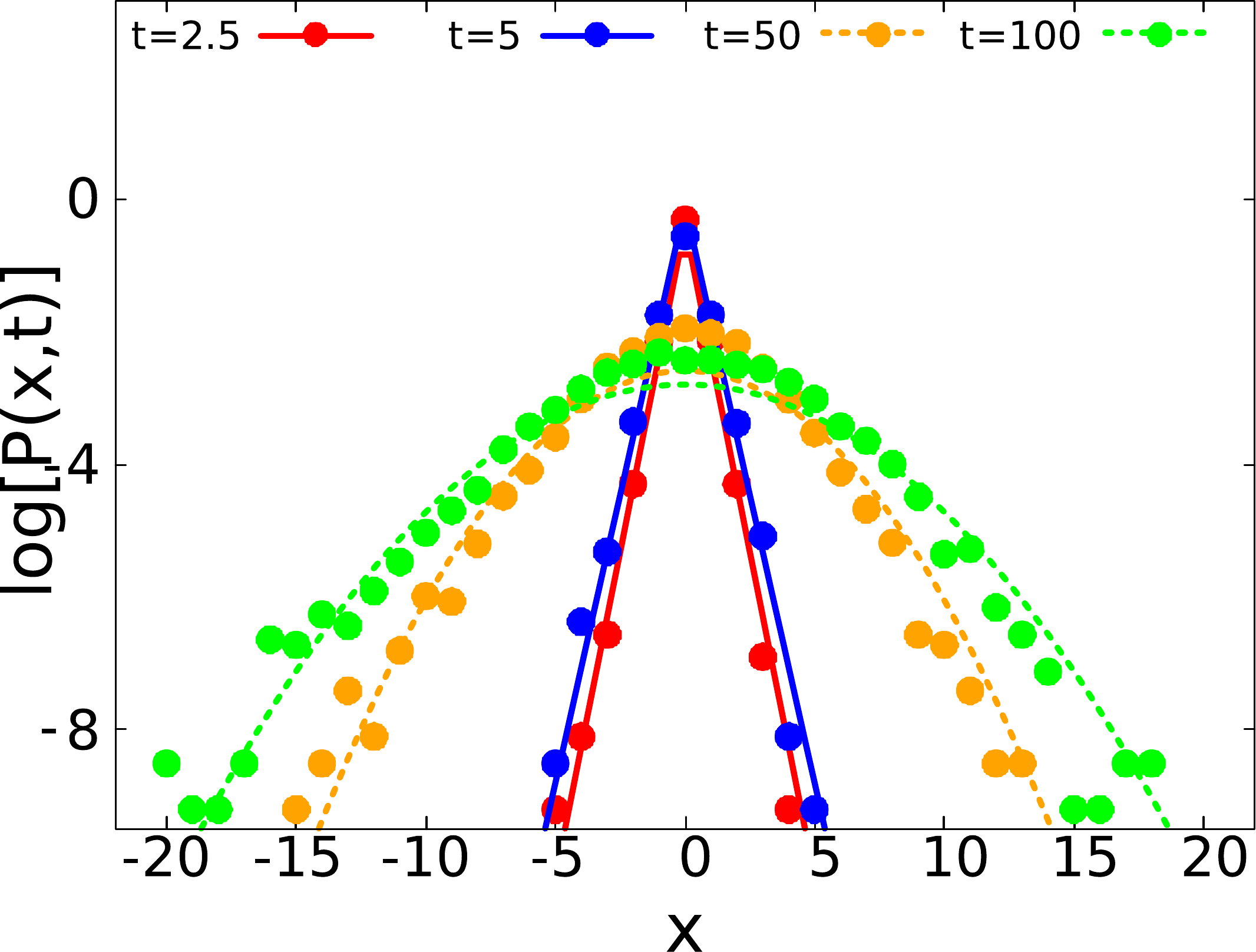}}} {} 
\mbox{\resizebox*{5.5cm}{4.0cm}{\includegraphics{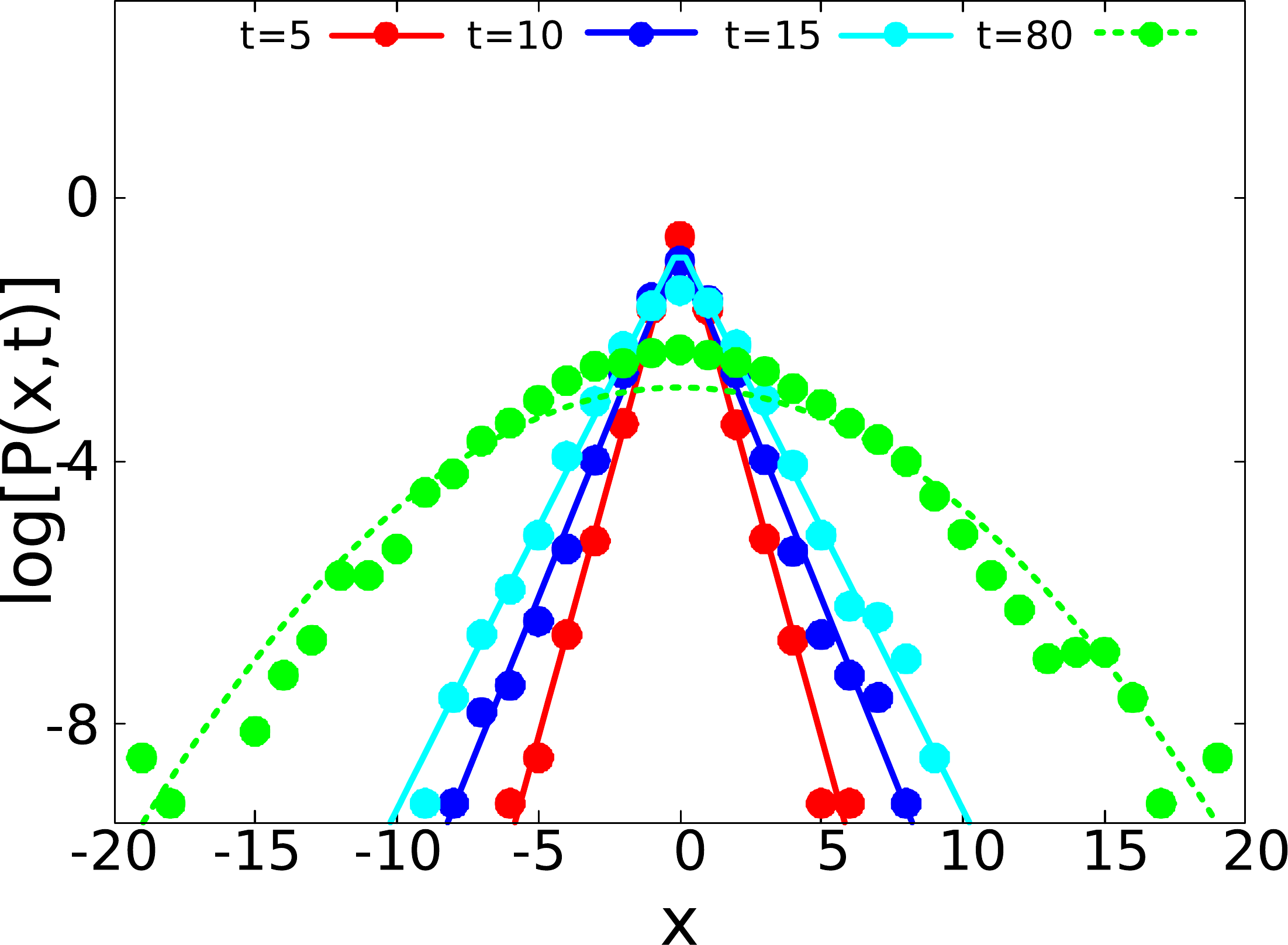}}} {} 
\end{center}
  \vspace{-6mm}
\caption{{\protect\footnotesize { (Color online)  Top:  Comparison between $P(N)$ obtained by the simulations of the Hitchhiker with Rouse dynamics but with different fragmentation mechanisms: equal binary breaking (red circles), random scission (purple circles), and chipping (black circles).  The fitting with Eq.~(3) with $\nu=1$ is shown in solid red line, the corresponding with Eq.~\eqref{eq:DNGSERB} in solid purple line, and $P(N)\sim N^{-\tau}$ in black  solid line. In the inset we show the same but in log-log scale, as we can see the power law model just explains $P(N)$ for the range of small sizes. Middle: Distribution of displacements $P(x,t)$ for simulations employing random scission. Bottom: $P(x,t)$ for simulations of a system with chipping. In both cases $P(x,t)$ exhibits an exponential decay in the short time limit, recovering Gaussian statistics in the long run. For both cases the particles were tracked by the SMT method. 
}}}
\vspace{0mm}
  \label{fig:pdfXRBC}
\end{figure}

In the top of Fig.~\ref{fig:pdfXRBC} we show in purple circles the molecule size distribution obtained from simulations of the Hitchhiker model with  random binary fragmentation  and Rouse dynamics. As we can see  $P(N)$ qualitatively has the same shape as in the case of equal binary breaking (red circles), but they differ in the peak height and the latter is slightly shifted to the right. Then is reasonable assuming that $P(N)$ still has the same functional form but now with different parameters. This is shown in  Fig.~\ref{fig:pdfXRBC} by the purple solid line,  which represents  the corresponding fitting with an inverse gamma distribution like model given by ~\cite{Mead2015} 
  
\begin{equation}\label{eq:DNGSERB}
P(N)=  CN^{-\beta}e^{-\frac{A}{N^{\delta}}}.
\end{equation}

The parameters relative to the powers of $N$ in Eq.~\eqref{eq:DNGSERB} are $\beta=2.83$ and $\delta=0.37$. In the middle panel of Fig.~\ref{fig:pdfXRBC} we observe that the displacements for short times follow a Laplace distribution, recovering Gaussian statistics in the long run. By changing the breaking mechanism for a more general one, we see that the distribution of sizes qualitatively remains equal and the displacements still exhibit an exponential decay in the short run.

\subsection{\label{sec:A92}Chipping}  

Finally we implement a single molecule breaking, also known as chipping mainly used in aggregation mass models studied in ~\cite{majum1998,raje2002}. In this case when a cluster of particles has a breaking event a monomer is ripped out from the aggregate and then it is placed randomly at one of the neighboring sites. So we have $N_{i}-1$ at the site of chipping and $N_{j}+1$ at $j=i+1$ or $j=i-1$. 

For the sake of comparison with the other cases mentioned above, we used chipping of particles in the Hitchhiker model with Rouse dynamics.  In the top panel of Fig.~\ref{fig:pdfXRBC} we show $P(N)$ in black circles,  as we can see this mechanism of breaking favors the existence of monomers, but also the creation of bigger size clusters. It is known from ~\cite{raje2002}, that for aggregation models with chipping, $P(N)$ follows a transition from exponential in the low density regime to a power  law distribution in the high density limit. In our case, since the density is given by $\rho=Total\,\ mass / L=1$, it is not clear which distribution should follow $P(N)$. In Fig.~\ref{fig:pdfXRBC} we show a power law fitting (solid black line) with the simulation data of the Hitchhiker with chipping, as we can see in the inset plot the power law $P(N)\sim N ^{-\tau}$ with $\tau=1.15$, fits well just for the range of small values of $N$. But in the large size regime this model does not describe anymore the behavior of $P(N)$. More importantly, the packet $P(x,t)$ exhibits  the now famous exponential decay, at least in the short time regime, see bottom panel of Fig.~\ref{fig:pdfXRBC}. As expected Gaussian statistics are recovered for the large time regime. 
\nocite{*}


\end{document}